# Strain Induced Strengthening of Soft Thermoplastic Polyurethanes Under Cyclic Deformation

*Giorgia Scetta, Jianzhu Ju, Nathan Selles, Patrick Heuillet, Matteo Ciccotti\* and Costantino Creton\**

Dr. G.S. Author 1, J.J. Author 2, Prof. M.C. Author 5, Prof C.C. Author 6

Sciences et Ingénierie de la Matière Molle, ESPCI Paris, Université PSL, CNRS, Sorbonne Université, 75005, Paris, France.

matteo.ciccotti@espci.psl.eu, Costantino.creton@espci.psl.eu

Dr. N.S. Author 4, Dr. P.H. Author 5

Laboratoire de Recherches et de Contrôle du Caoutchouc et des Plastiques, 60, Rue Auber 94408 Vitry-sur-Seine, France

**Abstract**

We investigate the cyclic mechanical behavior in uniaxial tension of three different commercial thermoplastic polyurethane elastomers (TPU) often considered as a sustainable replacement for common filled elastomers. All TPU have similar hard segment contents and linear moduli but sensibly different large strain properties as shown by X-Ray analysis. Despite these differences, we found a stiffening effect after conditioning in step cyclic loading which greatly differs from the common softening (also referred as Mullins effect) observed in chemically crosslinked filled rubbers. We propose that this self-reinforcement is related to the fragmentation of hard domains, naturally present in TPU, in smaller but more numerous sub-units that may act as new physical crosslinking points. The proposed stiffening mechanism is not dissimilar to the strain-induced crystallization observed in stretched natural rubber, but it presents a persistent nature. In particular, it may cause a local reinforcement where an inhomogeneous strain field is present, as is the case of a crack propagating in cyclic fatigue, providing a potential explanation for the well-known toughness and wear resistance of TPU.

## 1    Introduction

Soft TPU are well known for their outstanding combination of reversible elasticity, abrasion resistance and easy processability. They are multiblock copolymers characterized by alternating soft segments (SS) and hard segments (HS) forming a two-phase microstructure where the soft phase is the majority. The microphase separation is driven by the ability of the hard segments to form inter and intra-chain hydrogen bonds between carbonyl and amine groups developing therefore small and stiff lamellae domains surrounded by the soft phase. These hard domains (HD) have a typical size of 5-30 nm and are stiffer than soft segment domains (SD) acting both as nanoscale fillers and physical crosslinks [1], [2]. The versatile processing, recyclability and comparable reversible elasticity make TPU a serious competitor to replace elastomers in a number of technical applications despite the higher material cost.

Unsurprisingly, several studies on TPU focused on the correlation between their composition and mechanical properties at small and large strain [1], [3]–[5] with the target to further explore the field of applications of this class of materials. In particular, the presence of a mechanical behavior with the same characteristics as those of the Mullins effect classically observed in filled rubbers [6] (higher energy losses and stress softening during the first loading-unloading cycle), was reported in several TPU with different composition of hard and soft segments [3], [7], [8].

The Mullins effect was originally detected in filled (or crystallizing) chemically crosslinked rubbers [9] and despite its relevance for final rubber properties, its origin is still debated and may depend on the detailed structure of the material. Nevertheless, it is generally accepted, that the first cycle hysteresis comes in particular from structural rearrangements of filler aggregates [6]. In the case of unfilled TPU, the change in mechanical behavior after the application of strain is due to structural rearrangements of the multiphase structure. This mechanical effect has been often qualitatively investigated and compared to that observed in filled rubbers [2], [3], [8], [10], but to the best of our knowledge, has not yet been discussed quantitatively and in detail with a consistent methodology. Given the relevance of cyclic loading for the durability of materials and the renewed push to replace permanently crosslinked elastomers with thermoplastic reprocessable alternatives, we felt that the question of the evolution of the structure and mechanical properties of TPU with cyclic loading should be revisited.

Recently, Merckel and co-workers [11] proposed an easy methodology to quantify and compare the contribution of the structural changes to mechanical damage in filled elastomers after the application of a given strain. They proposed to use two damage parameters: one to account for the reduction in linear modulus and the other to account for the change in the onset of strain hardening in large strain. As they point out, these two parameters overlap for filled elastomers, and suggest that the same type of structural modification affects small strain stiffness and strain hardening [11]. However, mechanical damage is inherently related to the structural modifications induced by deformation which is a material-specific property and may be fundamentally different between TPU and filled crosslinked rubbers.

Using the above-mentioned methodology, we carried out here a quantitative evaluation of the structural changes occurring under cyclic loading in TPU and more importantly of the changes in mechanical properties induced by this cyclic loading. In order to be representative, we carried out all experiments with three commercially available TPU of the Elastollan™ series kindly provided by BASF AG.

Finally, within the framework of the possible substitution of conventional filled rubbers with TPU we also compared the mechanical behavior of our soft TPU with that of a typical crosslinked elastomer made from a random copolymer of styrene-butadiene (SBR) filled with reinforcing carbon black, emphasizing similarities between these two classes of materials in terms of small strain elasticity, and differences in structural evolution and large strain behavior.

## 2 Experimental results

### 2.1 Uniaxial tensile test

The different large strain behavior for all three TPU: TPU_XTAL, TPU_HARD and TPU_SOFT, can be appreciated in **Figure 1** (a), reporting the engineering stress-stretch curves in uniaxial tension at different temperatures. All TPU are characterized by a high extensibility at failure. The linear regime is only observed for a few percent of deformation (λ < 1.2) where we calculated the Young modulus E **Table 1** for 23 and 60°C). Above this limit, all three materials initially show a strain softening regime. Then TPU_XTAL and TPU_HARD present a marked strain hardening that is less intense in TPU_SOFT. Only in case of TPU_XTAL, as we reported in another work[12], the strain hardening is accompanied by the formation of a new crystalline phase (strain induced crystallization), which is partially retained in unloaded samples. The effect of increasing the temperature is mainly that of reducing the stress at break $\sigma_b$ for all materials, and in the case of TPU_SOFT also to reduce its maximum extensibility. Interestingly this reduction of strain at break with temperature was already observed for multiblock thermoplastic elastomers (TPE) based on poly-butadiene terephthalate (PBT) as hard segment as reported by Aime [13].

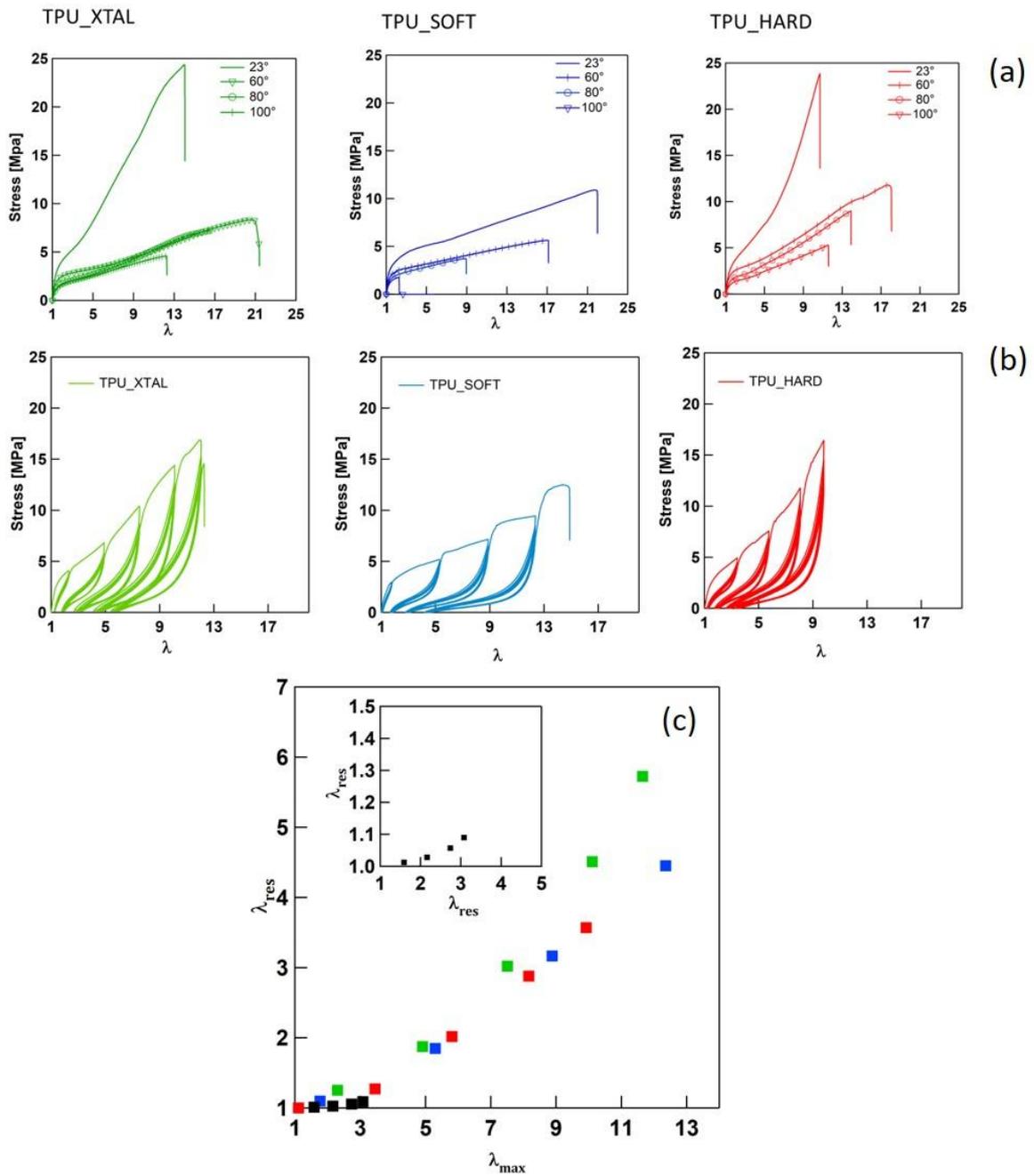

**Figure 1** Uniaxial stress-stretch curves at different temperatures (a) and cyclic stretch-stress curve at 23°C (b) for three TPU. (c) Residual vs. maximum applied deformation during the last cycle for: TPU_XTAL ■ TPU_SOFT ■ TPU_HARD ■ SBR ■

The stress-stretch curves of uniaxial step-strain cyclic test for all TPU at 23°C are reported in

**Figure 1** (b). All curves show similar characteristics:

- Large hysteresis between the first load-unload cycle.

- Very pronounced softening after the first loading, and recovery of the monotonic test behavior envelope only when the material is stretched to higher values than those previously applied (Mullins effect).
- Marked residual deformation $\lambda_{res}$ after unloading.

This large residual deformation $\lambda_{res}$ is consistent with literature data on TPU [3], [4] and can be attributed to the absence of chemical crosslinks as well as to the plastic deformation of the hard domains following the application of large strains. Interestingly, we found that for the same value of maximum applied stretch $\lambda_{max}$, all TPUs present a similar fraction of residual stretch $\lambda_{res}$ (**Figure 1** (c)). In contrast, in SBR the fraction of residual stretch is considerably lower for all values of tested maximum stretch compared to TPU because the chemical crosslinking prevents plastic deformation.

| Name | E @ 23°C [MPa] | E @ 60°C [MPa] |
| --- | --- | --- |
| TPU_XTAL | 8.7 ± 0.1 | 5.0 ± 0.7 |
| TPU_SOFT | 7.8 ± 0.1 | 5.8 ± 0.1 |
| TPU_HARD | 7.3 ± 0.1 | 5.7 ± 0.1 |

**Table 1** Young modulus for all TPU at 23 and 60°C

### 2.2 Structural investigations

WAXS 1D intensity profile are shown in **Figure 2** (a) for all three soft TPU. The absence of any crystalline reflection in TPU_XTAL and TPU_HARD indicates a completely amorphous hard phase in the pristine material. The crystalline peaks (indicated by the arrows) in TPU_SOFT are compatible with the crystalline structure of PBT as reported in literature [14], [15].

2D SAXS images and the corresponding integrated profiles for pristine and unloaded samples (previously strained for several cycles at $\lambda_{max}$= 2.25) are shown in **Figure 2** (b,c,d) for TPU_XTAL, TPU_HARD and TPU_SOFT respectively. All curves are characterized by a maximum in the intensity, due to microphase separation in TPU, which can be associated to the average distance between hard-domains (or long period L) [16] . The data of the long period L calculated from Bragg's law in both pristine and unloaded samples are reported in **Table 2**. The lower value of L for strained samples was already observed in similar multiphase systems [17]–[20] and was generally associated to a fragmentation of HD into smaller units as sketched in **Figure 2** (e).

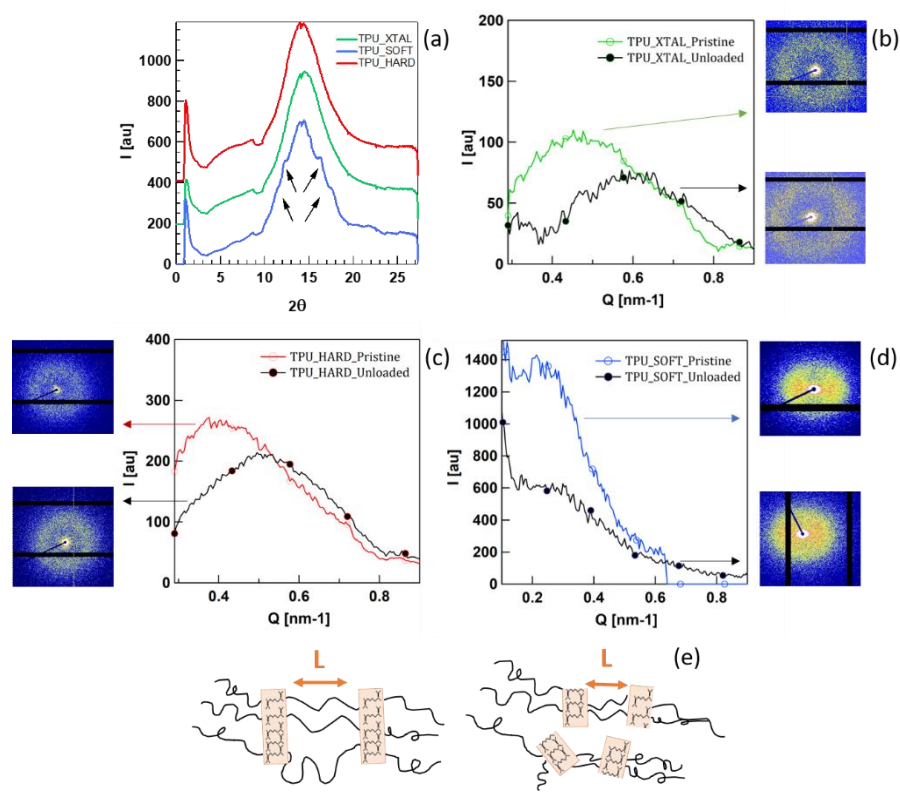

**Figure 2** (a) WAXS pattern for pristine TPU_XTAL, TPU_SOFT and TPU_HARD (data were vertically shifted for the sake of readability). (b) 2D SAXS Pattern and 1D integrated profile for pristine and unloaded TPU_XTAL(b), TPU_HARD (c), TPU_SOFT (d). Schematic illustrating the aggregation of HS in HD and definition of L (long period between hard domain) before and after loading (e).

Additionally, TPU_SOFT has an elliptical SAXS pattern indicating the presence of a preferential orientation in the HD (probably induced by injection molding) that contrasts with the random orientation of HD in the other two TPU (circular pattern).

We adopted the lamellar model proposed by Strobl and Schneider [21], which requires random orientation of HD, in order to evaluate some structural parameters in TPU_XTAL and TPU_HARD. Both materials present the same calculated values of thickness of hard domains (C) and hard segment volume fraction ϕ (**Table 2**) suggesting a very similar microstructure between TPU_XTAL and TPU_HARD. On the other hand, the higher value of L and the semi-crystalline character of HD in TPU_SOFT, indicates the presence of some major structural difference compared to TPU_XTAL and TPU_HARD.

| Name | L (Pristine) [nm] | L (Unloaded) [nm] | C [nm] | ϕ |
|---|---|---|---|---|
| TPU_XTAL | 13 | 10 | 2.9 | 0.24 |
| TPU_SOFT | 30 | 26 | - | - |
| TPU_HARD | 16 | 12 | 2.9 | 0.24 |

**Table 2** Long period L between HD for pristine and unloaded samples (according to Bragg's law). Domains thickness and hard segment volume fraction for pristine TPU_XTAL and TPU_HARD calculate using the lamellar model.

### 2.3 Damage analysis in cyclic loading

We used the approach proposed by Merckel et al. [11] to estimate the damage in both unfilled TPU and filled SBR in terms of large strain damage ($D_{ls}$) and small strain damage ($D_{ss}$) as detailed in the following. First, we expressed the data of cyclic loading experiments in terms of true stress vs. Hencky strain as reported in **Figure 3** (a) for TPU_HARD and SBR. We would like to stress that in cyclic deformation of TPU, the engineering stress-strain representation (**Figure 1** (b)), that seems to suggest a cyclic softening, is misleading. The engineering representation of strain in fact, only compares the final state with the initial state and, in the case of materials with high residual strain it introduces a non-negligible bias on the measurement. Contrarily, the Hencky strain accounts for all the incremental steps of deformation (considering the value of the sample length just before each strain increment).

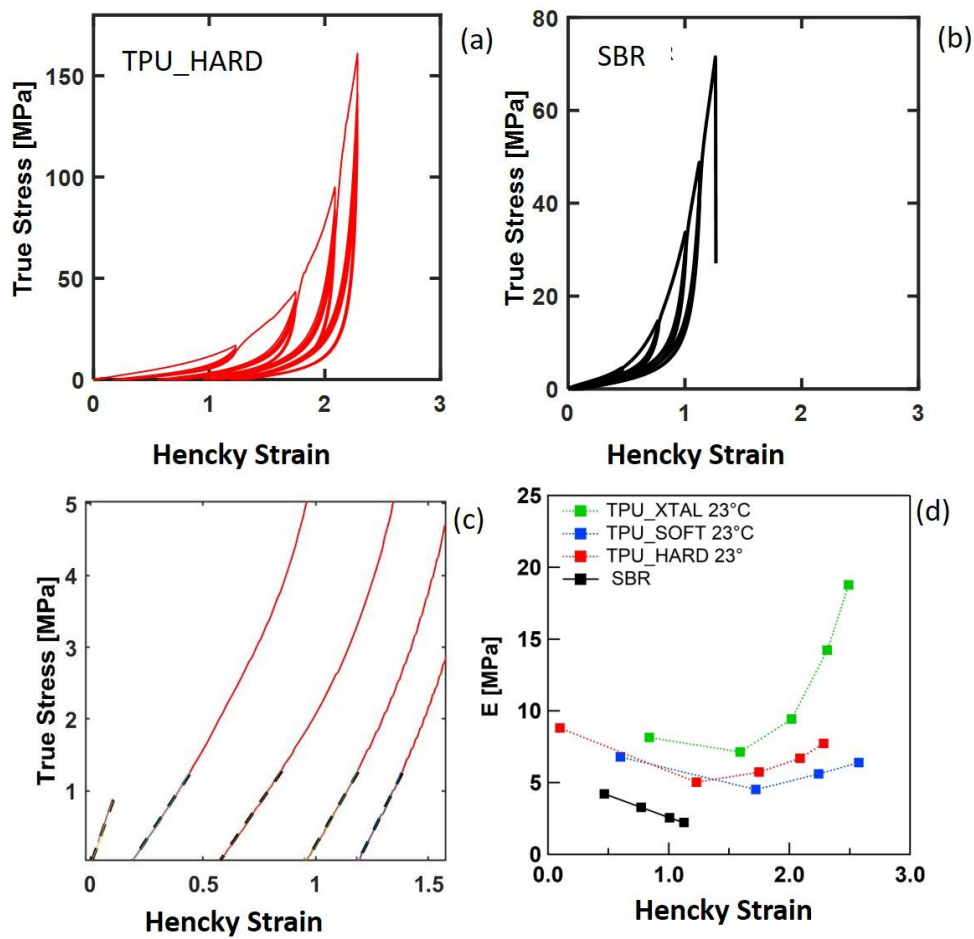

**Figure 3** True stress-Hencky strain representation of cyclic experiment for TPU_HARD (a) and filled SBR (b). Example of linear fitting for the calculation of $E$ for TPU_HARD (c). $E$ vs. Hencky strain for all TPU and for SBR (d).

The value of $D_{ss}$ for each increasing step of maximum strain $h_k$, is obtained from the ratio between the linear modulus $E_k$ for the cycles of step $k$ and that of the pristine material ($E_0$) as follows:

$$D_{ss} = 1 - \frac{E_k}{E_0} \qquad (1)$$

**Figure 3**(c) shows an example of linear fitting used to calculate *E,* and the values of the fitted modulus for all TPU and SBR are shown in **Figure 3**(d). The larger values of maximum Hencky strain at which the modulus was evaluated for TPU are justified by their higher maximum extensibility compared to SBR. While in the case of SBR, *E* always decreases with maximum applied strain, in all TPU *E* initially decreases and, for Hencky strains larger than 1.5 ($\lambda\sim4$), *E* increases again, reaching the pristine value or even larger values in the case of TPU_XTAL. A similar result was also found by Koerner [20]

The large strain damage, or $D_{ls}$, is associated to the onset of strain hardening in cyclic loading at different maximum strain and is defined as:

$$D_{ls} = 1 - \alpha \qquad (2)$$

where α is a re-scaling factor on the Hencky strain and it is obtained as follows:

1) Each unloading curve is shifted to the origin to compensate for the residual strain (**Figure 4**(a, c)).

2) A master-curve is built by performing a superposition fit using a least square minimization of each unloading curve using the first unloading curve as a reference (**Figure 4**(b, d)). This is mathematically equivalent to write: $h_{virgin} = \alpha * h_{shifted}$.

The first step is very important especially for materials such as TPU, which present a significant residual strain when unloaded. Shifting the stress-strain curve at the origin in fact, we obtain the same curve which a user, unaware of the previous strain history, would measure experimentally. Moreover, the shifted curves reported for TPU_HARD and SBR in **Figure**

4(a,c) demonstrate the different effect of maximum applied strain on the onset of strain hardening (approximatively indicated as $\lambda_{hard}$) between TPU_HARD and SBR. Unlike SBR, in TPU_HARD the onset of strain hardening in each unloading curve for the step $k+1$ appears at comparable or lower strain than step $k$. This has a strong influence on the previously described rescaling procedure as shown by the different values of the damage parameter $D_{ls}$ obtained for SBR and for all TPU.

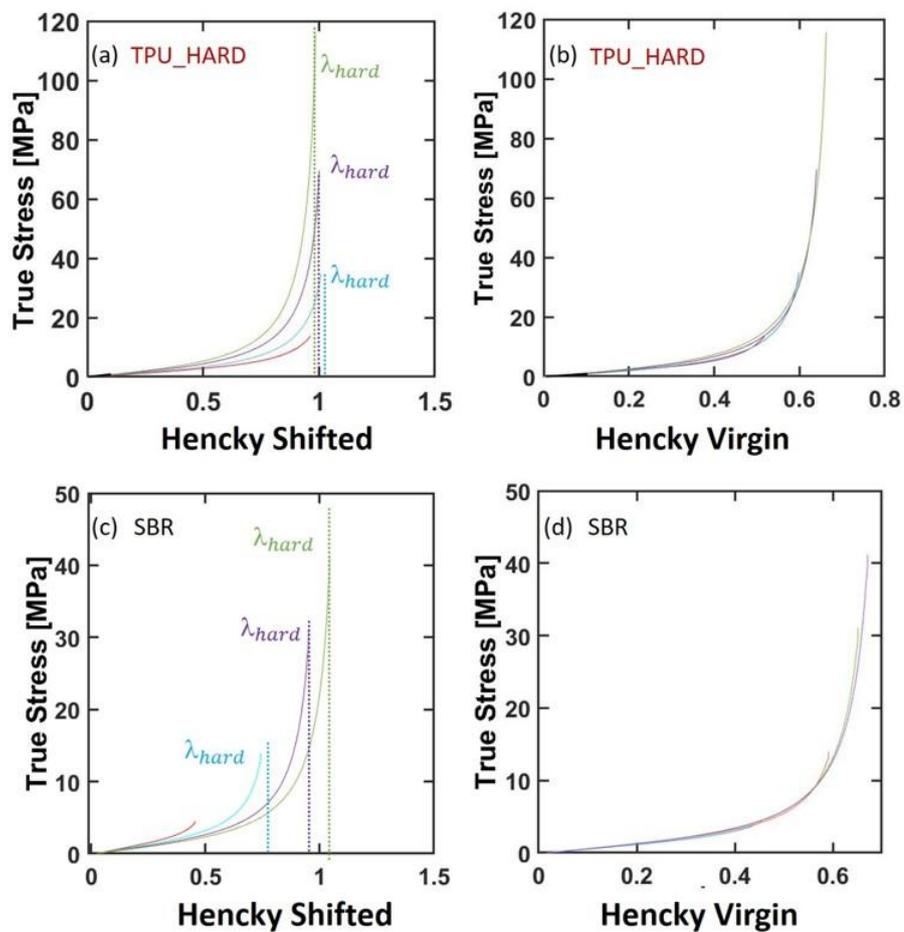

**Figure 4** Example of true stress vs. Hencky Shifted (a-c) and true stress vs. Hencky Virgin (b-d) respectively for TPU_HARD and SBR. The dotted lines represent an estimation of the onset of strain hardening and are used as a guide for the eyes.

**Figure 5** reports the calculated values of $D_{ss}$ and $D_{ls}$ for SBR at 23°C (a) and for all three TPU at 23 and 60°C (b, c). The trend is substantially different between SBR (where both $D_{ss}$ and $D_{ls}$ similarly increase with strain) and TPU. In all TPU, $D_{ss}$ and $D_{ls}$ have a dependence on the maximum Hencky strain that is qualitatively similar to the one we just discussed for the linear modulus. Both damage parameters first increase (softening) and then decrease (stiffening), eventually becoming negative in case of TPU_XTAL. At 60°C this trend is even more evident and all damage values are negative. Moreover, the damage in small and large strain are decoupled.

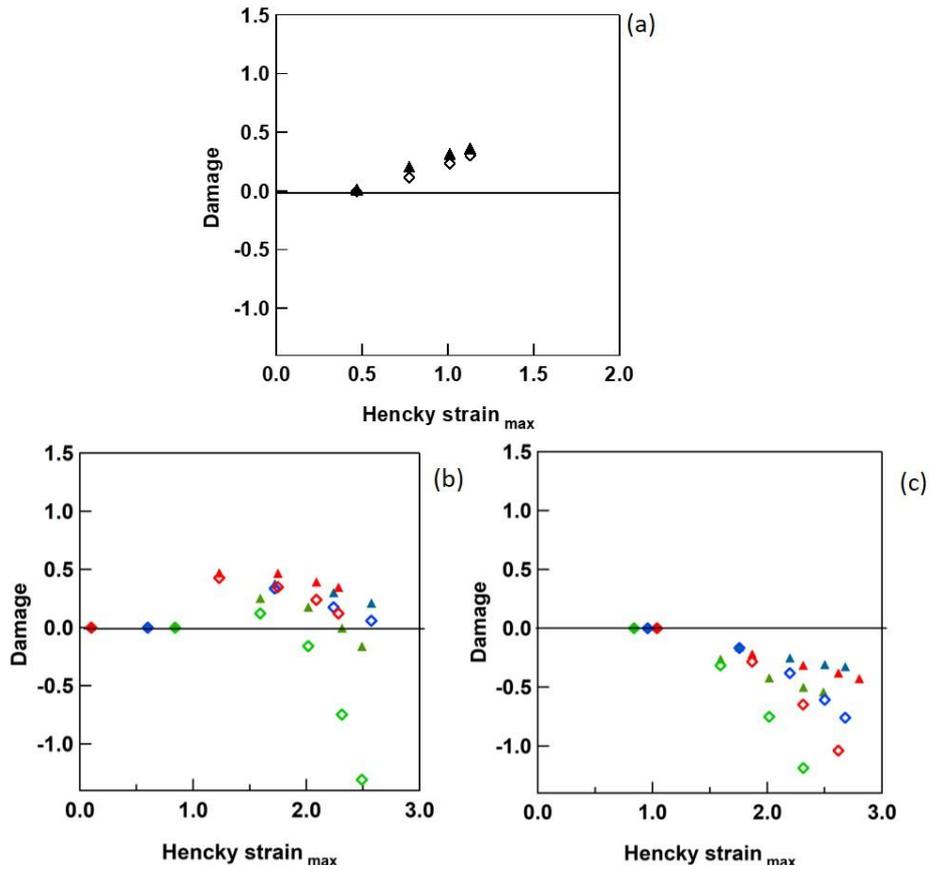

**Figure 5** Damage vs. Hencky strain calculated for SBR (a) and TPU at 23°C (b) and 60°C (c) . The symbols indicate $D_{ss}$ for: SBR: ▲, TPU_HARD: ▲, TPU_SOFT : ▲, TPU_XTAL: ▲ and $D_{ls}$ for SBR : ◊, TPU_HARD: ◊, TPU_SOFT: ◊, TPU_XTAL: ◊.

## 3   Discussion

### 3.1   Differences between TPU and SBR

In the case of SBR, the analysis of damage shows that both damage parameters ($D_{ls}$ and $D_{ss}$) nearly overlap and have a monotonically increasing dependence on the applied strain, meaning that the damage in the material increases with deformation. Merckel and coworkers [11] interpreted this as an indication that, in filled SBR, the application of a cyclic strain generates some kind of physical damage on the filled rubber that affects in a similar way both the small and large strain properties of the polymer itself. In other words, they argued that in SBR the change in the small and large strain mechanical response with applied strain have the same origin. On the other hand, although in all three TPU $D_{ls}$ and $D_{ss}$ have similar trends, they do not superimpose well and, above a threshold value of strain, they decrease with increasing maximum strain. In certain cases, $D_{ls}$ and $D_{ss}$ may assume even a negative value, especially at 60°C. Finally, it is important to underline that, the main difference between TPU and SBR, which leads to a different trend in the calculated damage, is the amount of residual deformation $\lambda_{res}$. In the case of SBR, $\lambda_{res}$ is mostly negligible while, in the case of TPU, $\lambda_{res}$ accounts for almost one third of $\lambda_{max}$ and has a great effect on the rescaling procedure.

## 3.2 Interpretation of the estimated damage in TPU

The decrease of $D_{ss}$ in TPU at large applied strains is associated with a permanent increase in the initial modulus $E$ with applied deformation for all samples (**Figure 3** (d)). In particular, TPU_XTAL, which can crystallize under strain at 23°C, shows the highest values of $E$ that eventually becomes even higher than in the pristine material, for large values of maximum applied strain, which results in a 'negative damage'. This inversion of trend is not typical of classical damage theories, where the damage parameter increases monotonically up to failure, as observed for SBR. The inversion of trend observed for TPU suggest that both estimators $D_{ls}$ and $D_{ss}$ reveal the competing effect of a strain induced damage mechanism with a strong strain induced stiffening mechanism that permanently affects both the initial modulus and the strain hardening properties of TPU. This unusual increase in the initial modulus with applied strain must depend on the specific multi-phase structure of each TPU that gets re-arranged with applied strain [16]. The reduction of the long period *L* in highly strained TPU was reported by several authors [17]–[19] and interpreted as a fragmentation of the original HD into smaller units. We believe that the breakdown of the original bigger hard domains into smaller, but probably more homogeneously dispersed and more numerous domains, has the same effect of increasing the density of physical crosslinking points, as schematically showed in **Figure 6** , thus causing an increase of the stiffness. In the case of TPU_XTAL, which crystallizes under strain, the increase in physical crosslinking (attributed to the fragmentation of hard domains) is enhanced by the intrinsic stiffness of the crystallites (that act as additional hard domains and crosslinking points), explaining the enhanced trend in *E(h)* for TPU_XTAL.

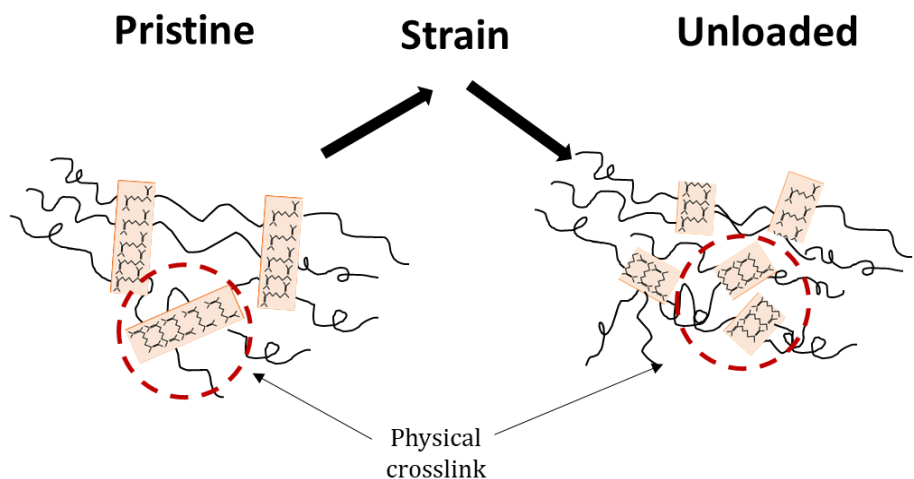

**Figure 6** Schematic illustration of HD restructuration with strain. The fragmentation of original HD in smaller but more numerous units provides additional physical crosslinking to the network.

## 4    Conclusions

We analyzed the cyclic mechanical behavior at small and large strain of three TPU which share a similar linear modulus (∼7-8 MPa at 23°C), but have different large strain properties, related to subtle differences in composition and microstructure. As common to most TPU, all of these multiblock copolymers self-organize into a microphase separated structure of hard domains that act as physical crosslinks into a softer matrix. This conveys to TPU their remarkable mechanical behavior, combining the classical elastomeric reversible elasticity with high extensibility, considerable strain hardening, damage softening limited to the first loading cycle (Mullins like), and contained plastic creep under large strain. We developed here a consistent methodology to assess these different components of the complex behavior of TPU, inspired from the work of Merckel and co-workers [11] on filled rubbers. This allowed us to identify the occurrence of a strain induced permanent stiffening in TPU, which is shown to increase progressively with applied strain in uniaxial cyclic loading, overcoming the damage softening effect that is common with filled rubbers.

A major focus of our work is the quantitative comparison of the different components of the mechanical behavior between three different commercially available TPU that were chosen to discriminate between role of macroscopic large strain behavior and the microscopic mechanisms involved in the structural modifications induced by cyclic loadings at large strain. While TPU_XTAL and TPU_HARD have similar fractions of HD and a remarkable strain-hardening, TPU_SOFT contains evidence of the presence of crystallized PBT and has a considerably less intense strain hardening before fracture. While the macroscopic large strain behavior of TPU_XTAL and TPU_HARD is quite similar, its microscopic origin for TPU_XTAL can be partly attributed to strain induced crystallization, that only occurs in this material. Despite their differences, we showed a surprising linear relationship between residual and maximum strain during cycles at different temperatures and all three TPU resulted in a pronounced change in mechanical properties after cyclic loading. However, the most remarkable and potentially impactful result of the present work lies in the discovery of strain induced stiffening in TPU with applied strain in cyclic loading. While the molecular origin of this peculiar behavior in TPU has not been clarified unambiguously yet, we propose that it originates from the fragmentation of original HD into smaller, but more numerous sub-units that may themselves act as additional physical crosslinking points [22]. This strain-dependent stiffening, can be compared to the strain-induced crystallization observed in stretched natural rubber. However, while NR melts when the strain is released, the SIC stiffening of TPU has a persistent nature and adds to the effect of HD fragmentation, which we propose as a more general mechanism of permanent strain stiffening in TPU. We believe that this peculiar strain induced permanent strain stiffening plays a major role in determining the remarkable resistance of TPU to cyclic fatigue, that we investigated in a companion paper [23] , due to the enhanced local stiffening of the crack tip region where the strains are locally concentrated by the presence of a notch.

## 5   Materials and methods

### 5.1   Materials

The three TPU used in this work are polyester based polyurethanes of the Elastollan™ series with commercial names EC 60 A 10P, LP9 277 10 and 565 A 12P and were kindly provided by BASF. We labelled them TPU_HARD, TPU_SOFT and TPU_XTAL respectively, to underline the difference in their large strain behavior as will be discussed in the experimental part. TPU_SOFT also differs from the others because it contains a small percentage of crystallized poly-butadiene terephthalate (PBT), while the other two are completely amorphous. Their glass transition temperature $T_g$ measured by differential scanning calorimetry (DSC at 10°/min) is reported in **Table 3.**

| Sample | TPU_HARD | TPU_SOFT | TPU_XTAL |
|---|---|---|---|
| $T_g$ a 10°/min °C | -50°C | -48°C | -34°C |

**Table 3** Glass transition temperature for all TPU measured by DSC

TPU specimens were injected by the Laboratoire de Recherche et Contrôle des Caoutchoucs et Plastiques (LRCCP) into 2 mm thick large square-plate, from which tensile dog-bone samples (cross section of 2x4 mm) were cut. The temperatures used in the injection molding procedure are summarized in the appendix for all samples.

The SBR rubber is filled with carbon black (CB) and all data concerning that rubber comes from the work of Mzabi et al. [24], [25]. The unvulcanised SBR has a mass $M_w$ of 120 kg/mol and a polydispersity of 1.94 and was provided by Michelin. Its styrene content is 15 wt% and the glass transition temperature $T_g$ measured by differential scanning calorimetry (DSC at 10°/min) is -48°C. The detailed composition is reported in Table 4 as provided by Michelin. All samples were prepared, molded and cured by Michelin. For tensile tests, samples were cut in a dog-bone shape and cross section of 2x4 mm and loaded with a strain rate of 4 s$^{-1}$.

|         | 20CB_19XL |
|---------|-----------|
| SBR     | 100       |
| N347    | 5         |
| 6PPD    | 1         |
| Struktol| 3         |
| CBS     | 1.5       |
| Sulfur  | 1.5       |
| $\phi$  | 0.03      |
| $\nu$   | $8.1 \cdot 10^{-5}$ |

**Table 4** SBR composition in PHR. Filler content ($\phi$) and crosslinking density ($\nu$) are reported in volumetric fraction (data from [24]). Note that N347 is a type of carbon black, Sulfur is a crosslinking agent, Struktol™ and N-Cyclohexyl-2-benzothiazole sulfenamide (CBS) are accelerators to vulcanize the rubber, and N-(1,3-dimethylbutyl)-N'-phenyl-p-phenylenediamine (6PPD) is an anti-oxidant

### 5.2  Step-Cycle Tests

The dog-bone shaped samples were strongly fixed between mechanical clamps since TPU are very tough. An optical system was used to measure the local stretch in the gauge area of the sample and to check the absence of slippage from the clamps during the test. The samples were strained in uniaxial conditions at the stretch rate of $\dot{\lambda}$ = 4 s$^{-1}$. The elongation was performed in a stepwise mode: 5 or 10 cycles were performed for each increasing value of maximum applied stretch $\lambda_k$ for SBR and TPU respectively. The stress was reduced to σ = 0 between two successive steps in order to prevent buckling. Strain ε, stretch λ, Hencky strain h, stress σ and true stress T are defined as below.

$$\varepsilon = \frac{l-l_0}{l_0} \qquad \lambda = \frac{l}{l_0} \qquad \sigma = \frac{F}{A_0} \qquad T = \sigma \cdot (1 + \varepsilon) \qquad h = \int_{l_0}^{l} d\varepsilon = \ln(\lambda)$$

$l_0$ and $l$ indicate the initial length and instantaneous length respectively, $A_0$ the initial cross section area and $F$ the measured force.

## 5.3 Structural study

Information on the physical structure of nanodomains was obtained by X-Ray analysis using the facilities of the materials science center of DSM, Netherlands. The beam wavelength was 0.154 nm. The 2D data were integrated using "FIT-2D" software [26]. All data were corrected by subtracting background scattering and circularly integrated to obtain 1D profile. SAXS data was expressed in terms of wave vector $q = \frac{4\pi \sin\theta}{\nu}$ where 2θ is the scattering angle and ν is the wavelength. 1D SAXS profile were fitted with Gaussian curves to determine the central position $q^*$ of the peak. According to Bragg's law the long period $L$ or inter-hard domain distance was evaluated as $L = \frac{2\pi}{q^*}$. The 2D SAXS profiles were obtained for all three TPU in relaxed conditions for two sets of samples: pristine and previously strained for several times at $\lambda_{max}= 2.25$.

An approach based on two-phase, lamellar morphology[21] and correlation function analysis was used to identify the crystalline layer thickness (C) and bulk volume crystallinity (ϕ) (graphically indicated in Figure 7) for those TPU presenting random orientation of HD.

The correlation function has the form:

$$K(x) = \int_0^\infty 4\pi q^2 I(q)\cos(2\pi qx)\, dq$$

(3)

This approach requires the extrapolation of the data at $q \to \infty$ and $q \to 0$. The data were extrapolated using Porod's law $I \approx q^4$ and Guinier's law: $I \approx A+Bq$ respectively.

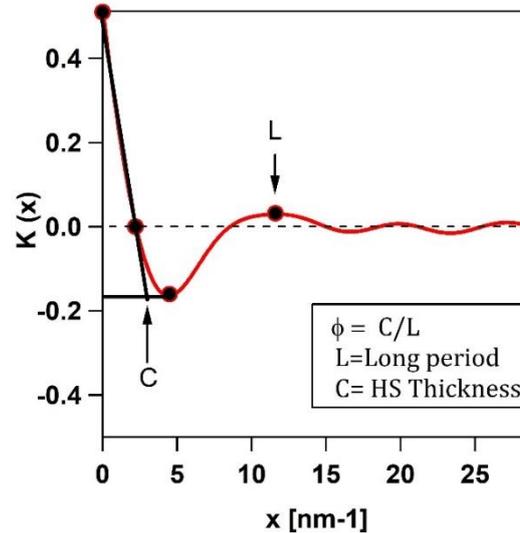

**Figure 7** Example of correlation function and interpretation of its features for TPU_HARD based on the analysis of Strobl and Schneider[21]

**Acknowledgements**

The PhD work of G. Scetta was jointly funded by the French ANRT and the LRCCP. We are grateful to Aude Belguise for her support with correlation function analysis. We are indebted to Dr. Matthias Gerst, Dr. Elke Marten and Mr. Stephan Dohmen from BASF AG for kindly providing the TPU samples. We thank Stephane Delaunay for injecting the samples. CC was partially funded by the European Research Council (ERC) under the European Union's Horizon 2020 research and innovation program under grant agreement AdG No 695351.

[17]  L. Wang, X. Dong, M. Huang, and D. Wang, "Transient microstructure in long alkane segment polyamide: Deformation mechanism and its temperature dependence," *Polymer (Guildf).*, vol. 97, no. May 2018, pp. 217–225, 2016, doi: 10.1016/j.polymer.2016.05.038.

[18]  R. S. Waletzko, L. S. T. James Korley, B. D. Pate, E. L. Thomas, and P. T. Hammond, "Role of increased crystallinity in deformation-induced structure of segmented thermoplastic polyurethane elastomers with PEO and PEO-PPO-PEO soft segments and HDI hard segments," *Macromolecules*, vol. 42, no. 6, pp. 2041–2053, 2009, doi: 10.1021/ma8022052.

[19]  K. Kojio, K. Matsuo, S. Motokucho, K. Yoshinaga, Y. Shimodaira, and K. Kimura, "Simultaneous small-angle X-ray scattering/wide-angle X-ray diffraction study of the microdomain structure of polyurethane elastomers during mechanical deformation," *Polym. J.*, vol. 43, no. 8, pp. 692–699, 2011, doi: 10.1038/pj.2011.48.

[20]  H. Koerner, J. J. Kelley, and R. A. Vaia, "Transient microstructure of low hard segment thermoplastic polyurethane under uniaxial deformation," *Macromolecules*, vol. 41, no. 13, pp. 4709–4716, 2008, doi: 10.1021/ma800306z.

[21]  G. R. Strobl and M. Schneider, "Direct Evaluation of the Electron Density Correlation Function of Partially Crystalline Polymers.," *J. Polym. Sci. Part A-2, Polym. Phys.*, vol. 18, no. 6, pp. 1343–1359, 1980, doi: 10.1002/pol.1980.180180614.

[22]  H. Zhang *et al.*, "Strain induced nanocavitation and crystallization in natural rubber probed by real time small and wide angle X-ray scattering," *J. Polym. Sci. Part B Polym. Phys.*, vol. 51, no. 15, pp. 1125–1138, 2013, doi: 10.1002/polb.23313.

[23]  C. Scetta, G. Selles, N. Heuillet P. Ciccotti, M. Creton, "Cyclic fatigue failure of TPU using a crack propagation approach," " *Prep.*

[24]  S. Mzabi, "Caractérisation et analyse des mécanismes de fracture en fatigue des élastomères chargés," pp. 1–310, 2010.

[25]  S. Mzabi, D. Berghezan, S. Roux, F. Hild, and C. Creton, "A critical local energy release rate criterion for fatigue fracture of elastomers," *J. Polym. Sci. Part B Polym. Phys.*, vol. 49, no. 21, pp. 1518–1524, 2011, doi: 10.1002/polb.22338.

[26]  A. P. Hammersley, "'FIT2D,'" *2016*. [Online]. Available: http://www.esrf.eu/computing/scientific/FIT2D/.

**Table of Content**

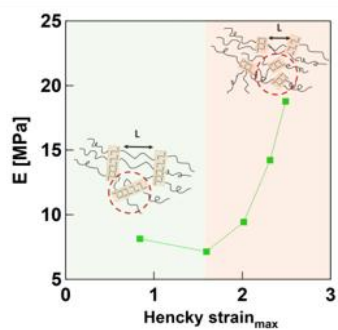

**Appendix**

The temperature used in the injection procedure is summarized in **A 1** and schematically showed in **A 2**

| Barrel / Name | Zone 1 (°C) | Zone 2(°C) | Zone 4(°C) | Nozzle (°C) | Mould (°C) |
|---|---|---|---|---|---|
| TPU_XTAL | 170 | 180 | 190 | 185 | 30 |
| TPU_SOFT | 190 | 200 | 205 | 200 | 30 |
| TPU_HARD | 165 | 170 | 175 | 170 | 30 |

**A 1** Barrel temperature profile for injection procedure. Zone 1 to 4 goes from the rear to the front of the barrel

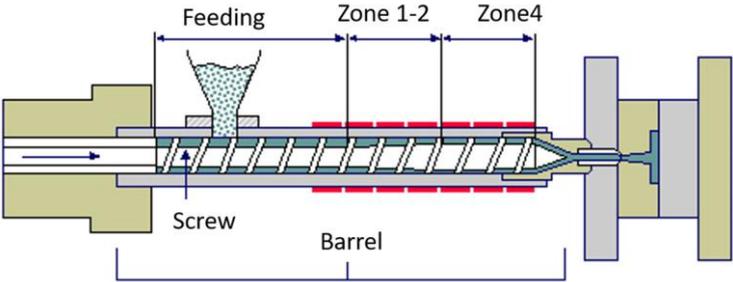

**A 2 S**chematic of injection process with the different zones of the barrel